\documentclass[12pt]{article}
\usepackage[utf8]{inputenc}
\usepackage{lineno}
\usepackage{comment}
\usepackage{siunitx}
\usepackage{subcaption}
\usepackage{authblk} 
\usepackage[style=numeric,sorting=none]{biblatex}
\usepackage{graphicx}
\usepackage{amsmath}
\usepackage{upgreek}
\usepackage{pdflscape}
\usepackage[english]{babel}
\usepackage{braket}
\newcommand{\SiN}{Si$_3$N$_4$ }

\begin{document}

\title{Cryogenic Graphene-Based Phase Modulators for Quantum Information Processing}

\author[1]{Leonard Barboza Navarro}
\author[1]{Maria Carolina Volpato}
\author[2,3]{Alisson Ronieri Cadore}
\author[1,*]{Pierre-Louis de Assis}
\affil[1]{\textit{Gleb Wataghin Institute of Physics, Universidade Estadual de Campinas, Campinas, Brazil.}}
\affil[2]{\textit{Brazilian Nanotechnology National Laboratory (LNNano), Brazilian Center for Research in Energy and Materials (CNPEM), Campinas 13083-200, São Paulo, Brazil.}}
\affil[3]{\textit{Programa de Pós-Graduação em Física, Instituto de Física, Universidade Federal de Mato Grosso, Cuiabá, 78060-900, Mato Grosso, Brasil.}}
\affil[* ]{Corresponding author: plouis@unicamp.br}

\date{} 

\maketitle

\begin{abstract} 

Electro-optic modulators are key components for photonic quantum computing, particularly in fully cryogenic integrated platforms where low loss and compactness are critical. We present a systematic theoretical investigation of compact dual-layer graphene (DSLG) electro-optic phase modulators integrated on silicon nitride (Si$_3$N$_4$) waveguides, with emphasis on cryogenic operation. By combining electromagnetic simulations with a physically consistent description of graphene conductivity based on the Kubo formalism, we analyze the interplay between electrostatic tuning, optical mode confinement, and material-dependent losses. We show that cryogenic operation enhances device performance by sharpening the Fermi–Dirac distribution, enabling access to the Pauli-blocking regime at lower Fermi levels and reducing the required modulation length. Through optimization of the waveguide geometry, dielectric spacer thickness and permittivity, and graphene quality, we identify regimes that simultaneously minimize insertion loss and device footprint under realistic voltage constraints. The optimized designs achieve near-pure phase modulation with insertion losses below \SI{0.3}{\dB} and modulation lengths below \SI{50}{\micro\meter} at 10 K, while maintaining GHz-scale bandwidths. These results provide quantitative design guidelines for low-loss, compact, cryogenic graphene phase modulators for scalable integrated quantum photonics.
\end{abstract}
\section{Introduction}
\noindent

Reconfigurable linear optical circuits are central to a wide range of integrated photonic technologies, including coherent optical communications, optical machine learning accelerators, and quantum information processing based on both discrete- and continuous-variable encodings \cite{shen2017deep,zhang2021optical,maring2024versatile,madsen2022quantum,larsen2025integrated}. In both classical and quantum applications, large operational bandwidths (typically above \SI{10}{GHz}) and low insertion loss are important figures of merit, as they directly impact data throughput, signal fidelity, and overall system scalability \cite{reed2010silicon,he2019high}. These metrics will be considered in this work when optimizing our designs.

Reconfigurable integrated photonic systems are commonly implemented using interferometric meshes composed of tunable beam splitters and phase shifters, where the Mach–Zehnder interferometer (MZI) serves as a fundamental unit cell. An integrated MZI can be composed of two 50:50 directional couplers and an internal phase shifter. When complemented by an additional external phase shifter, the device realizes a fully programmable beam splitter with independent control of the effective splitting ratio and output phase, enabling universal linear optical transformations connecting the input and output optical modes of the interferometer mesh \cite{Reck1994,Clements2016}. 

Phase modulators, unlike absorptive or amplitude-based modulators, ideally preserve photon number and therefore do not degrade detection probability, enabling low-loss linear transformations of quantum states. However, conventional implementations typically exhibit large device footprints, which, together with scaling requirements, constitute a major bottleneck for fully integrated photonic quantum information processing (QIP) systems. This challenge becomes even more pronounced in cryogenic architectures, where both single-photon sources and detectors are commonly operated at temperatures below \SI{10}{\kelvin} to suppress thermal noise and preserve quantum-state fidelity \cite{SinglePhoton2016}. At such temperatures, modulators based on materials such as LiNbO$_3$ and BaTiO$_3$ may exhibit partial or significant degradation of their electro-optic response \cite{Thiele2020,Chang2024,Qiao2007,Eltes2020}, further limiting their suitability for monolithic cryogenic photonic processors.

On the other hand, graphene, the archetypic two-dimensional material, exhibits enhanced electronic properties when cooled to cryogenic temperatures \cite{Lee2021}, especially carrier mobility. This positions graphene as an excellent candidate for use as the active medium in integrated modulators on silicon photonic platforms. Silicon alone is not an efficient electro-optic modulator because it lacks the Pockels effect \cite{Liu2023}. However, when integrated with graphene, its modulation performance can be significantly enhanced, enabling broadband operation over a wide range of wavelengths \cite{Liu2011}. While single layer graphene (SLG)-enabled refraction modulators on doped silicon have been demonstrated \cite{Wu2024,Shu2018}, the interaction between the guided optical mode and the graphene layer remains limited due to the strong field confinement within the silicon waveguide. High propagation losses imposed by heavy doping further limits their application as components in high performance classical circuits or in quantum circuits.

A more promising approach is to employ a dual single-layer graphene (DSLG) architecture, in which two graphene sheets form an integrated capacitor that enables phase modulation purely through electrostatic tuning, thereby eliminating the need for carrier injection in the waveguide core. This architecture allows the use of either silicon nitride (Si$_3$N$_4$) or undoped silicon as the guiding layer, both of which are compatible with CMOS fabrication and immune to carrier freeze-out effects at low temperatures \cite{Neamen2012}. Among these platforms, Si$_3$N$_4$ is particularly attractive due to its broad optical transparency window \cite{yang2018characteristic}, ultra-low propagation loss\cite{bose2024anneal}, good fabrication tolerance \cite{Fan2017}, and compatibility with deposition-based fabrication processes \cite{huang2006effect}. In addition, its moderate refractive index contrast enables efficient slot-waveguide geometries with strong electric-field confinement in the low-index region, which is highly beneficial for enhancing the overlap between the optical mode and the graphene layers.

Phase modulators based on Si$_{3}$N$_{4}$ with dual graphene layers have already been reported both theoretically \cite{Wu2024,Ji2020,Tiberi2025} and experimentally \cite{Wu2024,Mohsin2015,Watson2024}. Nevertheless, their optimization for low-power operation, minimized losses, and large modulation bandwidth under cryogenic conditions remains largely unexplored. In this context, the aim of this work is to investigate phase modulators based on graphene operating at low temperatures. We explored different waveguide geometries, dielectric materials used in dual single-layer graphene, and how the graphene quality and dimensions impact device performance at scale as a function of temperature.

\section{Modulation Principle} 
\noindent

The operating principle of graphene phase modulators is a change in optical conductivity caused by a change in the Fermi level of a monolayer placed in an electrical field \cite{}. When a gate voltage $V_G$ is applied across the symmetric DSLG capacitor configuration, the Fermi level $\mathrm{E_F}$ in each graphene layer can be tuned  according to \cite{Sorianello2015}
\begin{equation}
|V_G - V_\mathrm{Dirac}| =
\frac{e}{C_\mathrm{ox}} \frac{1}{\pi}
\left( \frac{\mathrm{E_F}}{\hbar v_F} \right)^2
+ 2 \frac{|\mathrm{E_F}|}{e}.
\label{eq:voltage_gate}
\end{equation}
Here $e$ is the electron charge, $\hbar$ is the reduced Planck constant, and $v_F$ is the Fermi velocity (considered in our work to be $\sim 9.5 \times 10^{5}\,\mathrm{m/s}$) \cite{}. 
$C_\mathrm{ox} = \varepsilon_r \varepsilon_0 / d_\mathrm{ox}$ is the geometric capacitance determined by the dielectric  relative permittivity $\varepsilon_r$, the vacuum permittivity $\varepsilon_0$, and the dielectric (typically an oxide, hence the subscript) thickness $d_\mathrm{ox}$.

The first term in Eq.~\ref{eq:voltage_gate} represents the electrostatic voltage drop across the dielectric, while the second term accounts for the electrochemical potential shift required to move the Fermi level in both graphene layers and is related to the finite density of states (quantum capacitance) of graphene. Depending on the position of the Fermi level, graphene may exhibit two distinct interaction regimes with a guided optical mode: an absorption regime and a transparency regime.

An optical field propagating inside the waveguide with photon energy $\hbar \omega$ interacts with the integrated DSLG capacitor. Due to the linear band structure of graphene, the condition $2E_F < \hbar\omega$ determines the interaction regime through Pauli blocking \cite{Wang2008}. When $E_F < \hbar\omega/2$, interband transitions from the valence to the conduction band are allowed, corresponding to the absorption regime exploited in graphene-based amplitude modulators. When $E_F > \hbar\omega/2$, interband transitions are suppressed due to Pauli blocking, leading to the transparency regime used for phase modulation. 

Tuning the Fermi level modifies the surface optical conductivity of graphene $\sigma(\omega)$, which determines the electromagnetic response of graphene to the guided optical mode. According to the Kubo formalism \cite{Falkovsky2008}, $\sigma(\omega)$ depends on the Fermi level, temperature $T$, scattering rate $\Gamma = 1/\tau$, and angular frequency $\omega$
\begin{equation}
\sigma(\omega) =
\frac{e^{2}\omega}{j\pi\hbar}
\left[
\int_{-\infty}^{+\infty} d\varepsilon\,
\frac{|\varepsilon|}{\omega^{2}}
\frac{d f_{0}(\varepsilon)}{d\varepsilon}
-
\int_{0}^{+\infty} d\varepsilon\,
\frac{f_{0}(-\varepsilon)-f_{0}(\varepsilon)}
{(\omega + j\Gamma)^{2} - 4\varepsilon^{2}}
\right],
\label{eq:surface_optical_conductivity}
\end{equation}
where $f_{0}(\varepsilon)=\left(1+\exp\left[(\varepsilon-E_F)/k_{B}T\right]\right)^{-1}$ is the Fermi--Dirac distribution. The parameters $\Gamma$ and $\tau$ in Eq.~\ref{eq:surface_optical_conductivity} represent the carrier scattering rate and relaxation time, respectively, accounting for momentum relaxation due to phonons, impurities, and structural defects. These parameters, together with the carrier mobility $\mu = e \tau v_F^2 / E_F$, are commonly used to characterize the material quality. Thus, the electrical tuning of the Fermi level through an applied gate voltage, together with the interaction between graphene charge carriers and the guided optical field, constitutes the two fundamental physical processes underlying electro-optic modulation in graphene-based devices.

In electromagnetic simulations, graphene is typically modeled as an infinitesimally thin conductive sheet whose surface conductivity modifies the boundary conditions of the electromagnetic field \cite{COMSOLGraphene2022,ANSYSGrapheneConductivity}. Therefore, the presence of graphene integrated on the waveguide modifies the complex effective refractive index $n_\mathrm{eff}$ of the guided mode. The real part $\mathrm{Re}(n_\mathrm{eff})$ governs phase propagation, while the imaginary part $\mathrm{Im}(n_\mathrm{eff})$ determines attenuation. Consequently, the accumulated phase shift and the modal power attenuation (MPA) are given, respectively, by
\begin{equation}
\Delta \phi = \frac{2 \pi}{\lambda} \Delta \mathrm{Re}\!\left(n_\mathrm{eff}\right) L, 
\quad
\mathrm{MPA} = \frac{40 \pi \log_{10} \mathrm{e}}{\lambda} \mathrm{Im}\!\left(n_\mathrm{eff}\right),
\label{eq:phase_shift_and_MPA}
\end{equation}
where MPA is expressed in decibels per unit length (dB/m), $\lambda$ is the free-space wavelength, $L$ is the modulation length, and $\mathrm{e}$ denotes the base of the natural logarithm.

Since both phase modulation and optical losses are directly encoded in \(n_{\mathrm{eff}}\), its real and imaginary components naturally provide a unified framework to define and compare standard performance metrics reported in the literature. Among the most established figures of merit are: (i) the modulation efficiency \(V_{\pi}L\) (\si{V \cm}), where \(V_{\pi}\) is the driving voltage required to induce a \(\pi\)-phase shift over a length \(L\); (ii) the insertion loss \(\mathrm{IL} = \mathrm{MPA} \times L\), which quantifies the total optical attenuation of the device; and (iii) the loss-efficiency figure of merit \(\mathrm{FOM} = V_{\pi}L \times \mathrm{MPA}\), which captures the intrinsic trade-off between modulation efficiency and propagation loss. In the context of quantum photonic circuits, where optical loss directly limits circuit depth and interference visibility, minimizing absolute insertion loss and device footprint becomes more critical than solely achieving record-low $V_{\pi}L$ values. In addition to electro-optic performance, device compactness is also of practical relevance in scalable quantum computing. Accordingly, we evaluate the minimum modulation length \(L_{\mathrm{min}}\) achievable before reaching the dielectric breakdown limit of the spacer material.

\section{Device Design and Optimization}

The simulations were performed using COMSOL Multiphysics. The parameters of the graphene capacitor were systematically varied, including graphene quality, operating temperature, dielectric material, and dielectric thickness, in order to evaluate their impact on modulation efficiency and optical losses. Finally, the electrical response of the phase modulator was also analyzed. We investigated a ridge waveguide (RWG) with a width of \SI{800}{\nano\meter} and a height of \SI{400}{\nano\meter}, featuring a \SiN core and a SiO$_2$ lower cladding, which supports only the fundamental TE mode. The separation between the edge of the core and the electrical contact was set to \SI{1.2}{\micro\meter} (see Supplementary Information S1 for details).

\subsection{Dual-Layer Graphene Capacitor Integration}

\begin{figure}[h!]
    \centering
    \includegraphics[width=1\columnwidth]{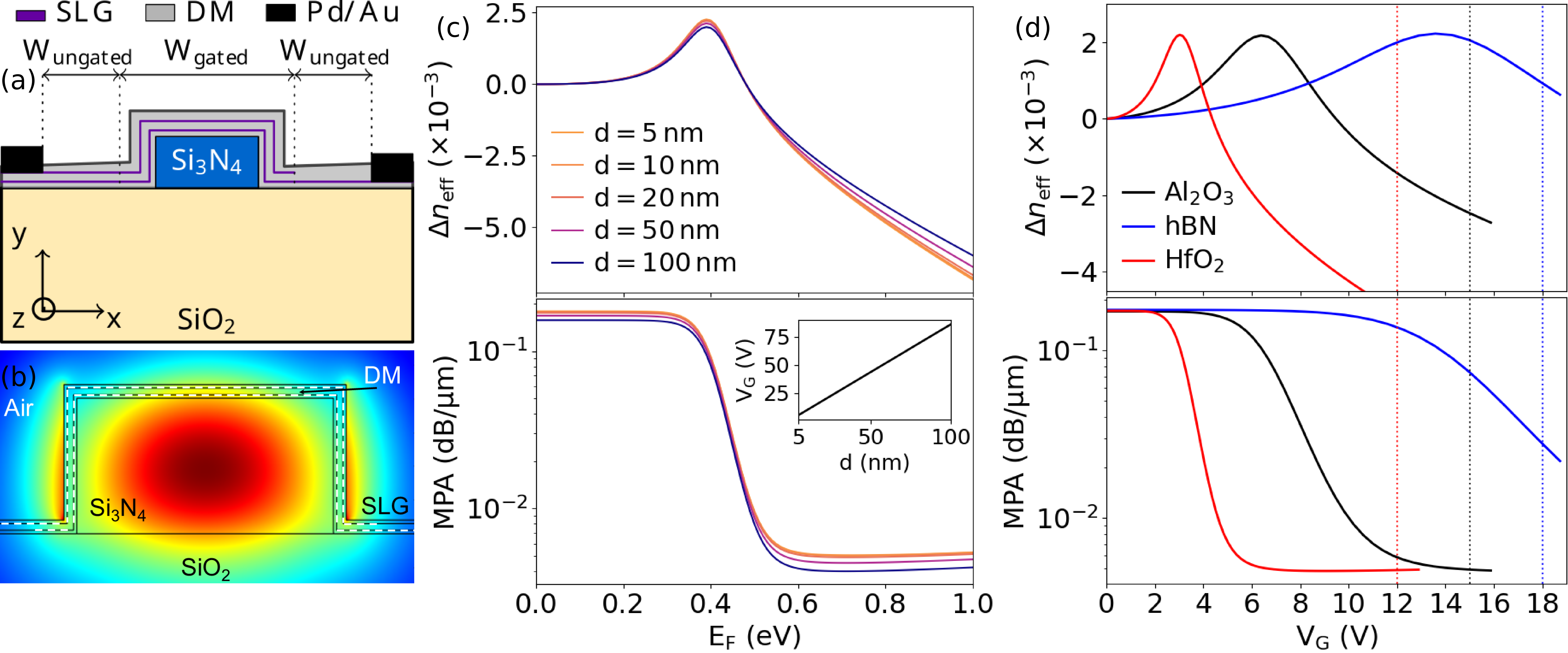}
    \captionsetup{width=1\columnwidth}
    \caption{
    (a) Cross-sectional schematic of the DSLG integration on the RWG. The SiO$_2$ buried oxide (BOX) layer is 2~$\mu$m thick, and the upper cladding is air. The core width and height were set to 800~nm and 400~nm, respectively. The gated region was set to approximately \SI{2}{\micro\meter}, while the ungated regions were approximately \SI{1}{\micro\meter} long. (b) Simulated |E| of the fundamental TE mode confined at \SI{1.55}{\micro\meter}. (c) Simulation $\Delta n_{\mathrm{eff}}$ and MPA for different hBN spacer thicknesses. The simulations were performed at $\lambda=\SI{1.55}{\micro\meter}$ and $T=\SI{300}{\kelvin}$. A carrier mobility of $\mu=\SI{10000}{\cm^2/\V\s}$ was assumed, corresponding to a relaxation time $\tau$ calculated at $E_F=\SI{0.11}{\electronvolt}$ according to $\mu = e\tau v_F^2/E_F$. The refractive index of hBN was taken as $n_{\mathrm{hBN}} = 2.0798$ \cite{Malitson1965}.
    Inset: Dependence of the required gate voltage on the hBN thickness for $\varepsilon_r=3.5$ at $E_F=\SI{0.45}{\electronvolt}$. (d) Simulated $\Delta n_{\mathrm{eff}}$ and MPA for different dielectric spacer materials. The simulations were performed under the same conditions as in (c). The refractive indices of Al$_2$O$_3$ ($\varepsilon_r = 8$) and HfO$_2$ ($\varepsilon_r = 20$) were taken as $n_{\mathrm{Al_2O_3}} = 1.7462$ \cite{Malitson1972Al2O3} and $n_{\mathrm{HfO_2}} = 1.8777$ \cite{AlKuhaili2004HfO2}, respectively.The dotted lines indicate the dielectric breakdown voltages for HfO$_2$ (\SI{12}{\V}, red), Al$_2$O$_3$ (\SI{15}{\V}, black), and hBN (\SI{18}{\V}, blue).
    }
    \label{fig:1}
\end{figure}

For the integration of the DSLG capacitor, comprising a SLG/dielectric/SLG heterostructure, as shown in the cross-section in Fig.~\ref{fig:1}(a), the optical surface conductivity $\sigma(\omega)$ was first computed numerically from Eq.~\ref{eq:surface_optical_conductivity}. All graphene sheets were assumed to be intrinsic, with no intentional chemical doping ($\mathrm{V_{Dirac}}$ = \SI{0}{\V}), such that the equilibrium Fermi level is taken as $E_F = 0$ \si{\eV}. In practice, charge transfer from the substrate can induce spatial charge-density fluctuations in graphene, commonly referred to as electron–hole puddles \cite{Droscher2010,Xia2009,SilvestreACSNano2013}. These fluctuations produce local variations of the Fermi level but preserve, on average, the charge neutrality condition over the optical length scale.

Similarly, charge transfer induced by the work-function mismatch between graphene and the metal contacts is typically confined to the immediate vicinity of the contact region and does not significantly affect the graphene regions located several micrometers away from the electrodes \cite{Giovannetti2008,Cadore2016,Pereira2019}. Both graphene layers were considered to be encapsulated in hexagonal boron nitride (hBN), which serves three main purposes: preventing carrier injection from the waveguide core or substrate into the graphene layers, protecting the graphene from environmental contamination, and providing an atomically flat interface that helps preserve the intrinsic electronic quality of graphene.

Due to the symmetry of the undoped DSLG configuration, the applied gate voltage induces equal but opposite shifts of the Fermi level in the two graphene sheets. As a result, both layers operate in the same optical regime while forming an electrostatically controlled capacitor structure.

\subsubsection{Dielectric Thickness Optimization}

Since the perpendicular electric field—and consequently the charge accumulation in the graphene layers—depends on the separation between the two SLGs, the thickness of the dielectric material (DM) spacer plays a critical role in the electro–optic response of the device. In addition, it determines the overlap between the optical mode and the SLGs, as shown in Fig.~\ref{fig:1}(b), thereby influencing the strength of the light–matter interaction. Therefore, a parametric study was performed to evaluate the influence of the dielectric thickness on the effective index variation and modal propagation attenuation. The results are shown in Fig.~\ref{fig:1}(c). 

From Fig.~\ref{fig:1}(c), three distinct regimes can be identified as a function of the Fermi level. For $E_F$ between \SI{0}{\electronvolt} and approximately \SI{0.3}{\electronvolt}, the graphene layers operate in the absorption regime, where interband transitions are allowed. Near \SI{0.4}{\electronvolt}, the onset of Pauli blocking produces a transition region characterized by a rapid change in both the effective index variation and the propagation losses. For Fermi levels above approximately \SI{0.55}{\electronvolt}, the system enters the transparency regime, where interband absorption is suppressed and phase modulation becomes dominant.

It can also be observed that increasing the separation between the upper and lower graphene layers reduces both $\partial \Delta n_{\mathrm{eff}}/\partial E_F$ and the MPA. This behavior arises because the upper graphene sheet moves farther away from the region of maximum optical field intensity of the guided mode, thereby reducing the modal overlap between the optical field and the graphene layers. To isolate the contribution of graphene to the overall losses, an additional simulation of the same structure without graphene was performed to confirm that the propagation losses are dominated by the graphene layers, even in the transparency regime, with intrinsic waveguide losses of approximately \SI{0.056}{\dB/\cm} due to the complex refractive index of Si$_3$N$_4$.

Since the optimization is focused on obtaining a design with low optical losses, larger dielectric thicknesses could be preferable. However, increasing the spacer thickness requires higher applied gate voltages, as shown in the inset of Fig.~\ref{fig:1}(c). Therefore, in order to achieve a suitable trade-off between phase modulation performance, optical losses and also electrical performance, the dielectric thickness was fixed at  \SI{20}{\nm}. Although the simulations consider Fermi levels up to \SI{1}{\electronvolt}, the dielectric breakdown field of hBN is approximately \SI{0.9}{\V/\nm}\cite{Mania2017}, which corresponds to a maximum voltage of about \SI{18}{\volt}. Under these conditions, the achievable Fermi level is limited to approximately $E_F \approx \SI{0.47}{\electronvolt}$, which prevents the device from fully reaching the transparency regime. For this reason, alternative dielectric materials must also be considered in order to enable higher electrostatic tuning of the Fermi level while maintaining low optical losses.

\subsubsection{Dielectric Spacer Material}

To achieve phase modulation in the transparency regime, we investigated the influence of CMOS-compatible dielectric spacer materials \cite{Yan2022} on the electro-optic response of the device, as shown in Fig.~\ref{fig:1}(d). Among the materials considered, HfO$_2$ emerges as the most suitable candidate due to its high relative permittivity, which enables access to the transparency regime at lower gate voltages. The dielectric material Al$_2$O$_3$ can achieve this regimen too. Although hBN exhibits slightly lower modal propagation attenuation (MPA) at $E_F = \SI{0.45}{\electronvolt}$ compared with HfO$_2$ and Al$_2$O$_3$, it requires significantly higher operating voltages. This increases the required $V_\pi$ and brings the device operation closer to the dielectric breakdown limit.

Furthermore, the operating voltage range for phase modulation using HfO$_2$ (approximately from \SI{6}{\V} to \SI{12}{\V}) enables larger values of $\Delta n_{\mathrm{eff}}$, allowing the realization of more compact modulators compared to those based on Al$_2$O$_3$ (approximately from \SI{11}{\V} to \SI{14}{\V}). Therefore, in the following simulations we consider an HfO$_2$ dielectric spacer, without excluding the use of Al$_2$O$_3$. The operating voltage is limited to \SI{11}{\V} to avoid operation close to the HfO$_2$ dielectric breakdown limit.

\subsubsection{Si$_3$N$_4$ capping layer}

\begin{figure}[h!]
    \centering
    \includegraphics[width=1\columnwidth]{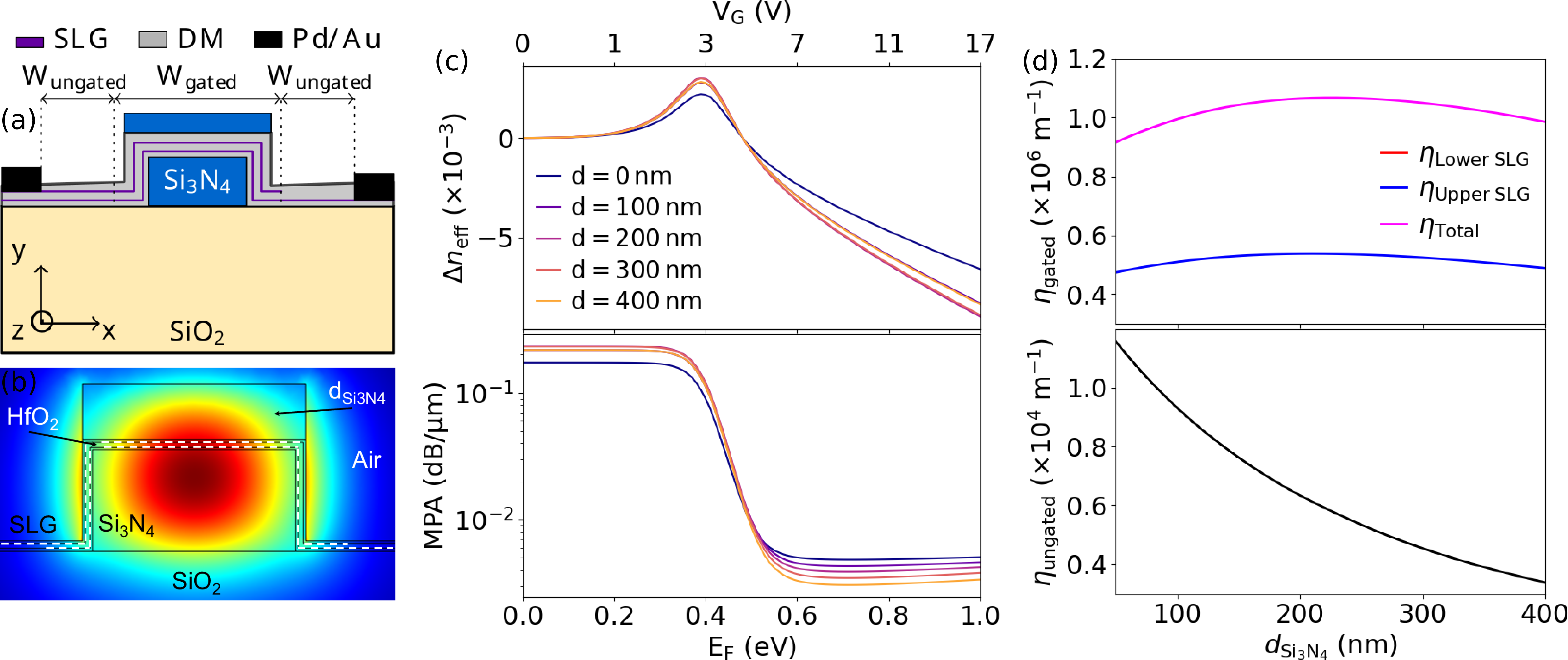}
    \captionsetup{width=1\columnwidth}
    \caption{
    (a) Cross-sectional schematic of the \SiN layer covering the DSLG integrated on the RWG. (b) Simulated |E| of the fundamental TE mode confined at \SI{1.55}{\micro\meter}. (c) $\Delta n_{\mathrm{eff}}$ and MPA for different values of the \SiN layer thickness. The simulations were performed under the same conditions as in Fig.~\ref{fig:1}(b-c), with \SI{20}{\nm} of HfO$_2$. (d) Modal overlap per unit length for the gated and ungated regions. Both were evaluated at $E_F = \SI{0.55}{\electronvolt}$; however, the same behavior is observed throughout the entire range (\SI{0}{\electronvolt} to \SI{1}{\electronvolt}). The overlap with the upper and lower graphene layers becomes nearly identical, as the region of highest optical field intensity also overlaps with the upper SLG.
    Inset: As the covering-layer thickness increases, the electric field $|E|$ is vertically redistributed into the upper \SiN layer. 
    }
    \label{fig:2}
\end{figure}

To improve the modulation efficiency, a \SiN layer is deposited on top of the DSLG integrated on the RWG, as shown in Fig.~\ref{fig:2}(a). This layer vertically redistributes the guided optical mode upward, bringing the region of highest optical field intensity closer to the DSLG capacitor (see Fig.~\ref{fig:2}(b)). Only the top of the core is coated, while the sidewalls remain uncoated to prevent the maximum of $|E_z|$ from shifting away from the vertical SLGs.

From Fig.~\ref{fig:2}(c), the highest phase modulation efficiency, $\partial \Delta n_{\mathrm{eff}}/\partial E_F$, is obtained for \SiN thicknesses of \SI{200}{\nm} and \SI{300}{\nm}. This behavior correlates with the modal overlap between the SLG and the optical mode (Fig.~\ref{fig:2}(d)), defined as $\eta=\int |E_{\parallel}|^{2}dl /\int \epsilon |E|^{2} dA$, which reaches its maximum in this thickness range due to the upward redistribution of the optical mode. For thicker \SiN layers, the overlap decreases as the optical mode increasingly extends into the \SiN, reducing the fraction of modal energy interacting with the graphene. Consequently, the optimal modulation efficiency is obtained for thicknesses between \SI{200}{\nm} and \SI{300}{\nm}. A similar trend is observed in the MPA in the absorption regime (\SIrange[]{0}{0.3}{\electronvolt}) due to the stronger modal overlap. In contrast, in the transparency regime the MPA decreases monotonically with increasing \SiN thickness. This reduction arises from the vertical expansion of the optical mode into the \SiN layer, which weakens the interaction between the evanescent field and the SLGs located in the ungated region, as quantified by $\eta_{\mathrm{ungated}}$, as shown in Fig.~\ref{fig:2}(d). Therefore, to achieve low optical losses while maintaining high modulation efficiency and a compact device footprint, the \SiN-layer thickness is chosen to be \SI{300}{\nm}.

\subsection{Impact of Graphene Material Parameters}

After optimizing the modal interaction between the graphene layers and the optical field, we now analyze the influence of the intrinsic graphene material parameters on the performance of the phase modulator, such as carrier mobility and its dependence on operating temperature.  

\subsubsection{Influence of Graphene Mobility}

\begin{figure}[h!]
    \centering
    \includegraphics[width=1\columnwidth]{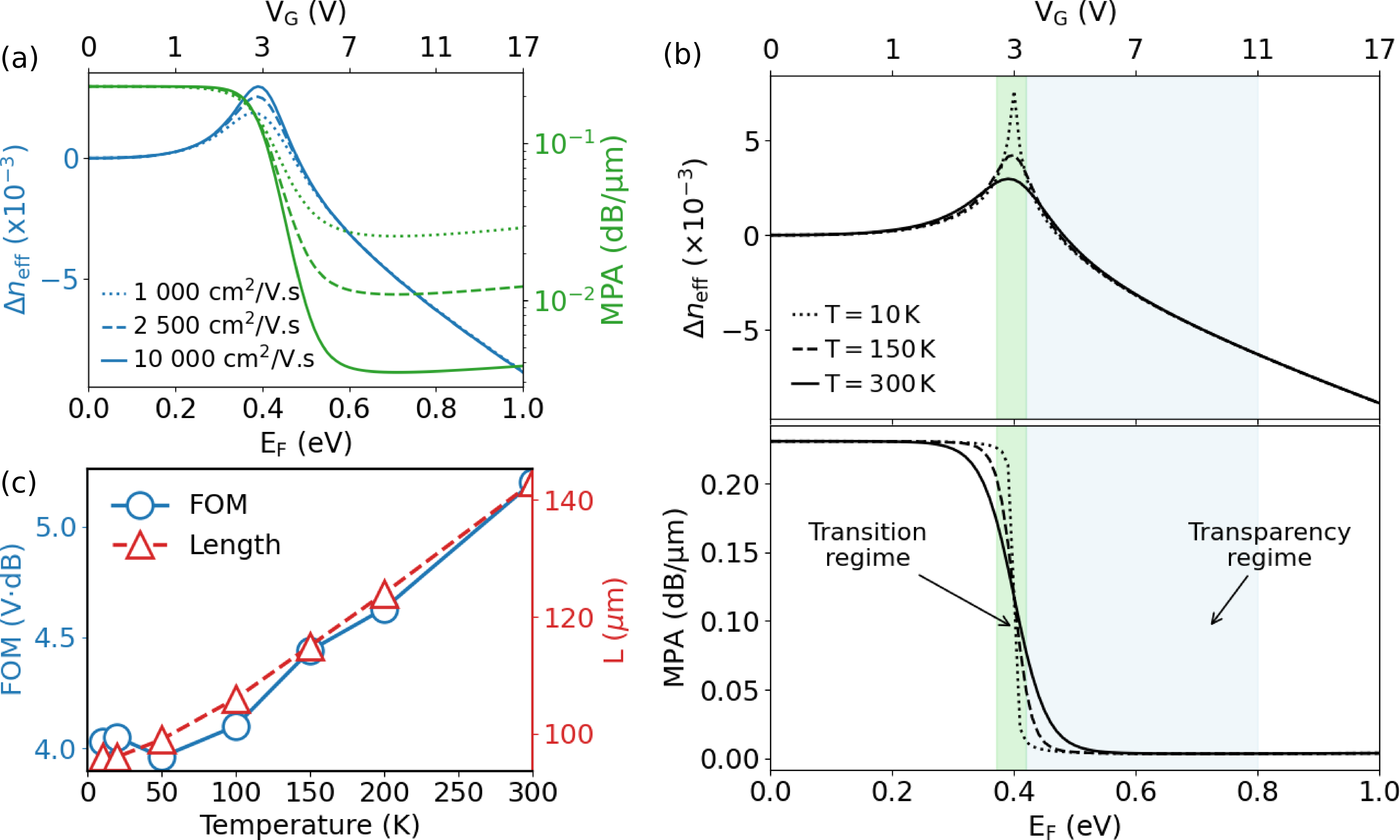}
    \captionsetup{width=1\columnwidth}
    \caption{
    (a) $\Delta n_{\mathrm{eff}}$ and MPA for different graphene carrier mobility values: low mobility ($\sim$ \SI{1000}{\cm^2/\V\s}), intermediate mobility ($\sim$ \SI{2500}{\cm^2/\V\s}), and high mobility ($\sim$ \SI{10000}{\cm^2/\V\s}). The simulations are performed using the same design parameters and conditions as in Fig.~\ref{fig:2}. (b) Simulations at different temperatures at \SI{10000}{\cm^2/\V\s}. As the temperature decreases, the transition regime becomes sharper due to the reduced thermal broadening of the carrier distribution. Moreover, the onset of the transparency regime shifts toward lower $E_F$. This enables access to operating regions where $\Delta n_{\mathrm{eff}}$ spans both positive and negative values, which can be exploited to significantly reduce the device footprint. The upper limit in the transparency regimen is to set at \SI{11}{\volt} to avoid the breakdown voltage. (c) Figure of merit (FOM) and minimum modulation length ($L_{\mathrm{min}}$) as a function of operating temperature, assuming a fixed carrier mobility of \SI{10000}{\centi\meter^2\per\volt\second}. The improvement in FOM at lower temperatures is primarily attributed to the reduction in modulation length, resulting in a more compact device footprint.
    }
    \label{fig:3}
\end{figure}

The carrier mobility $\mu$ characterizes the response of charge carriers to an external static electric field. In the presence of impurities, this response is degraded due to carrier scattering with impurity centers, leading to a reduced effective displacement \cite{Rezende2022,SilvestreACSNano2013}. This scattering process can be quantified through the relaxation time $\tau$, which represents the average time between successive collisions. Larger values of $\tau$ indicate a lower density of scattering centers and, consequently, higher graphene quality. From Fig.~\ref{fig:3}(a), it is observed that the carrier mobility has a limited impact on the phase modulation efficiency in the transparency regime, while it strongly affects the propagation losses. Specifically, the losses, evaluated at $E_F = $\SI{0.55}{\electronvolt}, decrease from \SI{0.0306}{\dB/\micro\meter} to \SI{0.0046}{\dB/\micro\meter} as the mobility increases from \SI{1000}{\centi\meter^2/\volt\second} to \SI{10000}{\centi\meter^2/\volt\second}, with an intermediate value of \SI{0.0134}{\dB/\micro\meter}. This range of values for the mobility is representative of typical CVD-grown graphene \cite{Banszerus2015,Gispare2021}.

These results highlight the critical importance of preserving graphene quality during synthesis, waveguide integration, and experimental characterization. For this reason, graphene sheets are usually encapsulated with hBN during fabrication in order to minimize impurity scattering and maintain high carrier mobility\cite{Viti2021}. Our design explicitly takes these experimental considerations into account.

It is worth noting that while the investigated mobility range (\SIrange{1000}{10000}{\cm^2/\V\s}) corresponds to typical CVD-grown graphene, mechanically exfoliated graphene can reach values up to $\sim$\SI{180000}{\cm^2/\V\s} \cite{Purdie2018} when fully encapsulated in hBN, potentially leading to even lower optical losses than those reported in Fig.~\ref{fig:3}(a).

\subsubsection{Temperature Dependence}
The transport properties of graphene, such as conductivity and carrier mobility, are strongly influenced by the operating temperature, as different scattering mechanisms dominate across distinct thermal regimes \cite{Heo2011}. At sufficiently high temperatures, electron–phonon scattering becomes the primary limiting factor, significantly reducing carrier mobility\cite{Purdie2018}. In contrast, at low temperatures, transport is mainly affected by potential fluctuations induced by charged impurities, which give rise to spatially inhomogeneous regions of electrons and holes, commonly referred to as electron–hole puddles \cite{Banszerus2015,SilvestreACSNano2013,Purdie2018}.

This disorder-induced scattering mechanism becomes particularly relevant when the amplitude of the potential fluctuations is comparable to or exceeds the Fermi level $E_F$, a condition typically associated with low carrier densities. For higher $E_F$ values, corresponding to increased carrier densities, the influence of these fluctuations is progressively suppressed, leading to a reduced impact on charge transport \cite{Heo2011}. Consequently, operating at low temperatures mitigates electron–phonon scattering and, in high-density regimes, minimizes disorder-induced effects, resulting in an enhanced mean free path and longer relaxation times, thereby improving the effective carrier mobility\cite{Purdie2018,Banszerus2016}.

In addition to these advantages, the carrier distribution around the Fermi level becomes sharper under cryogenic conditions due to the suppression of thermal activation. As a result, the Fermi–Dirac distribution approaches an ideal step-like profile, with electronic states being more distinctly occupied. As shown in Fig.~\ref{fig:3}(b), decreasing the temperature enables access to the transparency regime at lower Fermi levels (from $\sim$\SI{0.4971}{\electronvolt} at \SI{300}{\kelvin} to $\sim$\SI{0.4250}{\electronvolt} at \SI{10}{\kelvin}). This leads to wider transparency windows and larger $\Delta \mathrm{Re}(n_\mathrm{eff})$ and $d \Delta n_{\mathrm{eff}}/dE_F$, ultimately allowing for a reduced modulator footprint,

\section{Device Performance}
\subsection{Optical Performance}

The optical response is characterized by the changes in the complex effective refractive index induced by the DSLG capacitor. They are extracted from Fig.~\ref{fig:3}(b) at \SI{300}{\kelvin} and \SI{10}{\kelvin}. To achieve a phase modulation of $\Delta \phi = \pi$ under a single-arm (SA) modulation scheme, the minimum modulation length (active length) is calculated over the full transparency window. This range starts at \SI{4.8}{\volt} for \SI{300}{\kelvin} and at \SI{3.66}{\volt} for \SI{10}{\kelvin}. The upper limit of the transparency regime is determined by the cutoff voltage. Based on these values, the modulation efficiency $V_{\pi} \cdot L$ is evaluated. The key optical performance metrics, such as \(V_{\pi}L\), IL, and FOM, are summarized in Table~\ref{tab:opt_elec_metrics}.

\begin{landscape}
\begin{table}[ht]
\centering
\captionsetup{width=1\columnwidth}
\caption{Comparison of electro-optic modulators. IL: insertion loss. MZM: MZI modulator. SA: Single-arm MZI. PP: Push-pull MZI. FOM is defined as $V_{\pi}L \cdot \mathrm{MPA}$ when available.}
\resizebox{1.5\textwidth}{!}{
\begin{tabular}{c c c c c c c c c c c c}
\hline
Ref. & Material & Demo & Type & Device & Temp. [\si{\kelvin}] & {$\lambda$ [\si{\nm}]} & {$V_{\pi}L$ [V$\cdot$cm]} & {IL [\si{\dB}]} & {$L$ [\si{\um}]} & Speed [\si{GHz}]  & FOM \\
\hline

\multicolumn{12}{c}{\textbf{Graphene-based modulators}} \\
\hline
\cite{Ji2020} & DSLG & Theoretical & Phase & MZM & 300 & 1550 & 0.0125 & 0.64 & 80 & 48.3 & 1 \\
\cite{Watson2024} & DSLG & Exp. & Phase & MZM & 300 & 1550 & 0.3 & 5.6 & 75 & 24 & 3 \\
\cite{Wu2024} & DSLG & Exp. & Phase & MZM & 300 & 1550 & 0.0954 & 13 & 477 & 4.2 & 27.6 \\
\cite{Tiberi2025} & DSLG & Theoretical & Phase & WG & 300 & 3800 & 0.48 & $<$1 & 200 & 11 & N/A \\
\cite{Sorianello2015} & DSLG & Theoretical & Phase & MZM & 300 & 1550 & 0.16 & 0.6 & 500 & 30 & 1.92 \\
\cite{Shu2018} & SLG/Si & Theoretical & Phase & MZM & 300 & 1510--1600 & N/A & 2.8 & 10000 & N/A & N/A \\

\hline
\textbf{This work} & DSLG ($\mathrm{HfO_2}$) & Theoretical & Phase & MZM (SA) & 300 & 1550 & 0.0887 &  0.84 & 143 & 6.5 & 5.3 \\
\textbf{This work} & DSLG  ($\mathrm{HfO_2}$) & Theoretical & Phase & MZM (SA) & 10 & 1550 & 0.0704 & 0.55 & 96 & 6.5 & 4.0 \\
\textbf{This work} & DSLG ($\mathrm{HfO_2}$)  & Theoretical & Phase & MZM (PP) & 10 & 1550 & 0.0704 & 0.28 & 48 & 6.5 & 4.0 \\

\textbf{This work} & DSLG ($\mathrm{Al_2O_3}$) & Theoretical & Phase & MZM (SA) & 300 & 1550 & 0.1348 &  2.3 & 383 & 15.2 & 8.0 \\
\textbf{This work} & DSLG  ($\mathrm{Al_2O_3}$) & Theoretical & Phase & MZM (SA) & 10 & 1550 & 0.1004 & 0.94 & 164 & 15.2 & 5.8 \\
\textbf{This work} & DSLG ($\mathrm{Al_2O_3}$)  & Theoretical & Phase & MZM (PP) & 10 & 1550 & 0.1004 & 0.47 & 82 & 15.2 & 5.8 \\

\hline
\multicolumn{12}{c}{\textbf{Silicon-based modulators}} \\
\hline

\cite{Li2017SiliconTransmitter} & Si PN junction & Theoretical & Phase & MZM & 300 & 1550 & 1.5 & 6.7 & 2470 & N/A & 40.5 \\
\hline
\multicolumn{12}{c}{\textbf{Ferroelectric modulators}} \\
\hline

\cite{Zhang2021LNReview} & Thin-Film LiNbO$_3$ & Exp. & Phase & MZM & 300 & N/A & 1.4 & 0.5 & 2000 & $>$45& 0.4 \\
\cite{Eltes2020} & Thin-Film BaTiO$_3$ & Exp. & Phase & MZM & 4 & 1550 & 5 & $<$1 & 500 & 30 & N/A \\

\hline
\end{tabular}}
\label{tab:opt_elec_metrics}
\end{table}
\end{landscape}
The insertion loss is calculated at the bias voltage corresponding to $L_{\mathrm{min}}$, yielding approximately \SI{7.3}{\volt} at \SI{10}{\kelvin} and \SI{7.9}{\volt} at \SI{300}{\kelvin}. The figure of merit (FOM), which captures the trade-off between phase modulation efficiency and optical losses, is also evaluated at the corresponding bias voltage for each temperature.

The observed improvement in FOM is primarily attributed to the reduction in modulation length, as shown in Fig.~\ref{fig:3}(c). It is important to note that the simulations were performed as a function of temperature assuming a constant carrier mobility of \SI{10000}{\centi\meter^2\per\volt\second}. This assumption is conservative, as higher mobilities are typically reported at low temperatures, particularly for high-quality CVD and exfoliated graphene. An alternative configuration is the push-pull (PP) scheme, in which both arms of the Mach–Zehnder interferometer (MZI) contribute equally to the phase modulation by inducing phase shifts of $+\pi/2$ and $-\pi/2$. In this case, the required device length is reduced by a factor of two. However, the $V_{\pi} \times L$ product remains unchanged, since the driving voltage is effectively doubled due to the simultaneous modulation of both arms.

\subsection{Electrical Performance}

To evaluate the electrical performance of the proposed modulators, we adopt figures of merit commonly used in electro--optic telecommunications, enabling a direct comparison with state-of-the-art devices. Although these metrics are typically defined under classical, high-photon-flux operation, they are also relevant in quantum photonics, where fast and deterministic control of single-photon states is required. The energy consumption per bit, $E_{\mathrm{bit}} = C\,V_{pp}^{2}/2$, is determined by the capacitive charging of the device. In the present DSLG architecture, the equivalent capacitance $C$ arises from the series combination of the oxide capacitance and the graphene quantum capacitance. Under a symmetric drive configuration, the energy required to induce a $\pi$-phase shift scales with the square of the applied peak-to-peak voltage $V_{pp}$. In the push--pull scheme, where both interferometer arms are driven simultaneously, the total energy consumption increases by a factor of two due to the presence of two identical capacitive loads.

The modulation speed is fundamentally limited by the RC time constant of the device, which defines the electrical 3-dB bandwidth, $f_{\mathrm{3dB}} = 1/(2\pi R C)$. This bandwidth determines the maximum rate at which the graphene capacitor can be charged and discharged, and therefore sets the limit for dynamic phase modulation. In quantum photonic circuits, a sufficiently large bandwidth is essential to address individual photons in time-resolved operation without temporal overlap, thereby preserving coherence and interference visibility \cite{Slussarenko2019}. The total electrical resistance $R$ is governed by the graphene sheet resistance, defined as $\sigma_{\mathrm{d.c.}} = 1/R_{\mathrm{sheet}} = N e \mu$, which depends on the Fermi level and carrier mobility, as well as by the contact resistance at the metal-graphene interface. Importantly, the gated and ungated graphene regions exhibit different effective resistances due to their distinct carrier densities N. While the ungated regions are charge-neutral on average, spatial charge inhomogeneities (electron-hole puddles) provide a finite conductivity that dominates the dc transport response. In contrast, the gated region operates at elevated Fermi levels within the transparency regime, resulting in a lower sheet resistance.

\begin{figure}[h]
    \centering
    \includegraphics[width=0.5\textwidth]{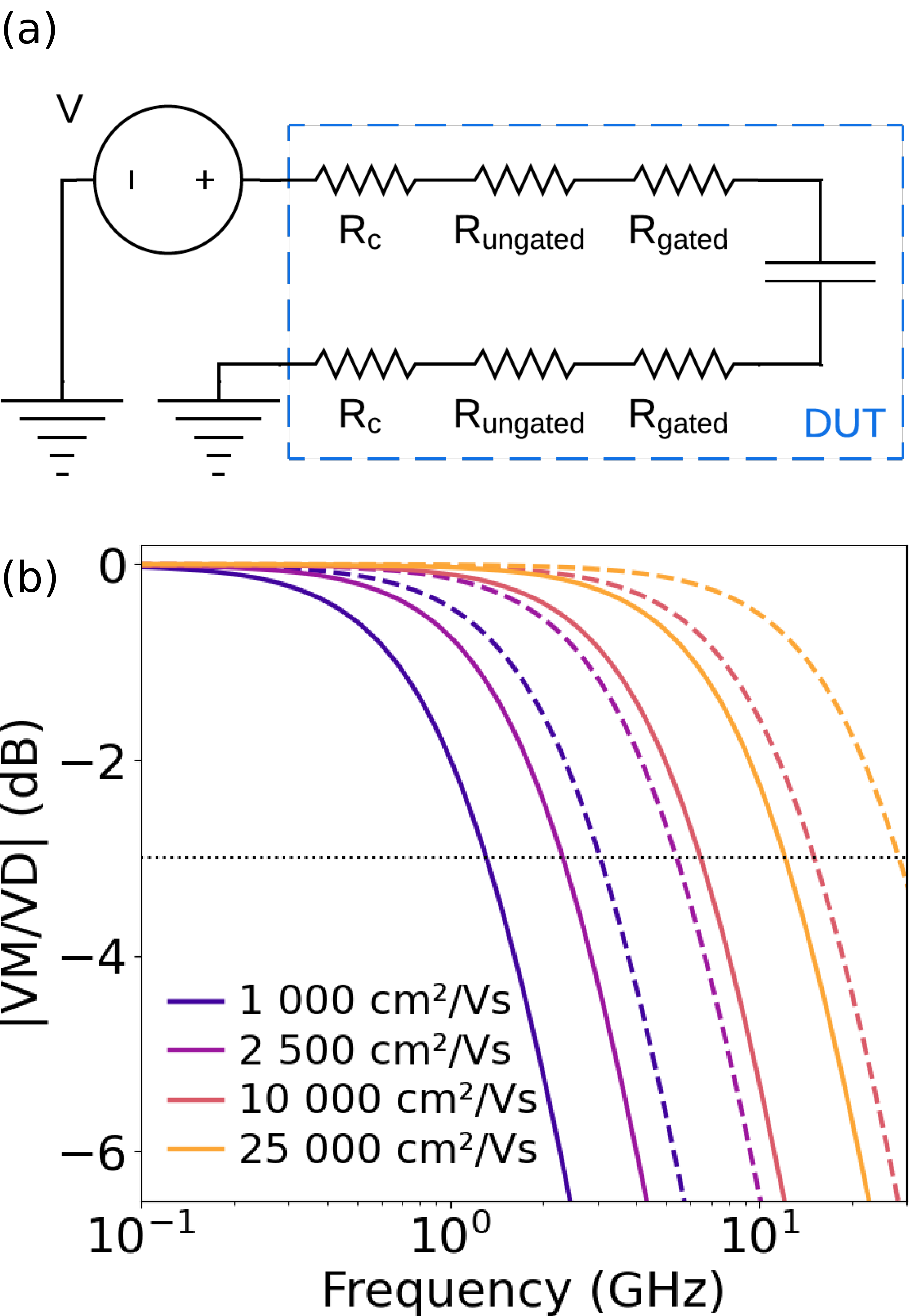}
    \captionsetup{width=1\columnwidth}
    \caption{
    (a) Equivalent electrical circuit of the DSLG-based waveguide phase modulator. 
    (b) Frequency response of the DSLG-based waveguide phase modulator for different carrier mobility values ($\mu$), assuming a contact resistance of $R_c = $\SI{150}{\ohm \um} \cite{Ji2020}. The graphene in the gated region ($\sim$ \SI{2}{\um}) is set to $E_F =$ \SI{0.55}{\eV}, while in the ungated region ($\sim$ \SI{1}{\um}) it is set to $E_F =$ \SI{0.11}{\eV}. Solid lines corresponds to \SI{20}{\nm} HfO$_2$ spacer ($\epsilon_r = 20$) with modulation length of \SI{96}{\um}.
    Dashed lines correspond to \SI{20}{\nm} Al$_2$O$_3$ spacer  ($\epsilon_r = 8$) with a modulation length of \SI{164}{\um}. }
    \label{fig:4}
\end{figure}

The overall electrical response of the device can be described using an equivalent circuit that includes contact resistances, ungated graphene sections, and the actively gated DSLG capacitor, as shown in Fig.~\ref{fig:4}(a). Within this framework, only the capacitive component, characterized by an impedance $Z_C(\omega) = 1/(j\omega C)$, directly contributes to the modulation of the optical phase, as it governs the charge accumulation responsible for tuning the Fermi level. However, the presence of series resistance reduces the fraction of the applied voltage $V_D$ that effectively drops across the capacitor $V_M$. As a result, the device behaves as a first-order low-pass filter, with a voltage transfer function of the form $V_M/V_D = 1/(1 + j\omega R C)$, where the modulation efficiency decreases at high frequencies due to incomplete charging of the graphene capacitor. For frequencies approaching the RC cutoff, the achievable phase shift is reduced, preventing the system from reaching a full $\pi$-modulation depth. The results are shown in Table~\ref{tab:opt_elec_metrics}.

Figure~\ref{fig:4}(b) shows a 3-dB bandwidth of approximately \SI{6.5}{\GHz} for a carrier mobility of \SI{10000}{\centi\meter^2\per\volt\second}, assuming a metal--graphene contact resistance of \SI{150}{\ohm \um}, as reported for  lateral contacts. The response is strongly influenced by the graphene quality, as higher mobility reduces the resistance in both the gated and ungated regions, thereby improving the effective modulation voltage. Another key limiting factor is the metal--graphene contact resistance. For the same mobility, the \SI{3}{\dB} bandwidth decreases to \SI{3.1}{\GHz} when $R_c =$ \SI{1000}{\ohm \um} is considered. To enhance the modulator speed, simulations were performed using a \SI{20}{\nm } Al$_2$O$_3$ layer with a mobility of \SI{10000}{\centi\meter^2\per\volt\second} and $R_c =$ \SI{150}{\ohm \um}, yielding a bandwidth of \SI{15.2}{\GHz}, and up to \SI{28.2}{\GHz} for \SI{25000}{\centi\meter^2\per\volt\second}, as shown in Figure~\ref{fig:4}(c). Therefore, selecting an ideal dielectric material, improving graphene quality, and optimizing contact engineering are essential to achieve high-speed modulators.

From Table~\ref{tab:opt_elec_metrics}, the proposed modulator exhibits bandwidths in the \si{GHz} range. Although these values are lower than those of established LiNbO$_3$ and BaTiO$_3$ platforms, as well as some previous theoretical works~\cite{Ji2020,Sorianello2015}, they remain several orders of magnitude higher than the typical repetition rates of deterministic single-photon sources, which are on the order of \SIrange[]{2}{10}{\MHz}~\cite{Slussarenko2019}. This limitation is particularly relevant for low-loss thermo-optic modulators, whose electrical performance is further degraded under cryogenic operation. Therefore, even for relatively low graphene mobilities, the proposed device provides sufficient temporal bandwidth to enable phase control of individual photons without limiting the source repetition rate.

Nevertheless, there are clear pathways to further increase the bandwidth. One approach is to employ exfoliated graphene, whose mobility can easily exceed \SI{25000}{\centi\meter^2/\volt\second}, and even higher values at cryogenic temperatures. Under these conditions, bandwidths of up to \SI{12.2}{\GHz} can be achieved for a contact resistance of $R_C =$  \SI{150}{\ohm \um}. However, the use of exfoliated graphene is inherently limited by its small available area, typically restricted to lateral dimensions on the order of a few tens of micrometers, which poses significant challenges for large-scale integration and wafer-level fabrication. Another strategy is to modify the gate dielectric, replacing HfO$_2$ ($\varepsilon_r = 20$) with Al$_2$O$_3$ ($\varepsilon_r = 8$). This substitution enables bandwidths approaching \SI{30}{\GHz} when combined with exfoliated graphene and edge-contact configurations. These results indicate that the electrical response is primarily limited by the capacitance of the active modulation region.

Notably, although the design based on Al$_2$O$_3$ results in a slightly longer device compared to HfO$_2$, the footprint remains in the sub-\si{\um} range for a push--pull configuration. This confirms the potential of the proposed architecture as a strong candidate for compact electro-optic modulators in both quantum photonics and telecommunications applications.

\section{Conclusion} 
\noindent 

In this work, we have presented a comprehensive theoretical study of dual-layer graphene electro-optic phase modulators integrated on Si$_3$N$_4$ waveguides, with particular emphasis on cryogenic operation. By combining electromagnetic simulations with a physics-based description of graphene conductivity, we established a consistent framework to evaluate the interplay between electrostatic tuning, optical confinement, and material-dependent losses.

Our results demonstrate that cryogenic operation provides a clear advantage for graphene-based modulators. The suppression of thermal broadening sharpens the Fermi–Dirac distribution, enabling access to the Pauli-blocking regime at lower Fermi levels. This effect increases the effective index modulation while maintaining low absorption, allowing a substantial reduction in device footprint without compromising optical performance.

Through systematic optimization of waveguide geometry, dielectric spacer thickness and material, and graphene quality, we identified the key design trade-offs governing device efficiency. In particular, high-permittivity dielectrics such as HfO$_2$ enable efficient electrostatic tuning at reduced voltages, while the use of Al$_2$O$_3$ further reduces the required device length due to improved field confinement and reduced charge disorder effects. At the same time, maintaining high graphene mobility remains essential to minimize losses in the transparency regime.

At \SI{10}{\kelvin}, the optimized devices achieve modulation lengths of $L =\SI{82}{\um}$ for HfO$_2$ spacers and $L =\SI{48}{\um}$ for Al$_2$O$_3$ spacers, corresponding to voltage–length products of $V{_\pi}L=\SI{0.10}{\volt\cm}$ and \SI{0.07}{\volt\cm}, respectively. These values are obtained while maintaining low insertion losses and GHz-range bandwidths, limited primarily by the RC response of the graphene capacitor. Although this bandwidth remains lower than that of state-of-the-art LiNbO$_3$ platforms, it is well above the repetition rates of typical single-photon sources, making it suitable for quantum photonic applications.

Overall, this work establishes practical design guidelines for compact, low-loss, and cryo-compatible graphene-based phase modulators, where cryogenic operation not only preserves but enhances electro-optic performance. These results position DSLG architectures as a scalable platform for integrated quantum photonic circuits, where minimizing optical loss, footprint, and power consumption is essential for high-fidelity quantum operations.

\section*{Acknowledgments}
L. B. N. and P.-L. de A. acknowledge the support of National Council for Scientific and Technological Development (CNPq), Process Number 409516/2022-8. A.R.C. acknowledges the support of CNPq Grants 301145/2025-3 and 408783/2024-9.

\printbibliography

@article{Slussarenko2019,
  author  = {Slussarenko, S. and Pryde, G. J.},
  title   = {Photonic quantum information processing: A concise review},
  journal = {Applied Physics Reviews},
  year    = {2019},
  volume  = {6},
  number  = {4},
  pages   = {041303},
  doi     = {10.1063/1.5115814}
}

@article{Reck1994,
  author  = {Reck, M. and Zeilinger, A. and Bernstein, H. J. and Bertani, P.},
  title   = {Experimental realization of any discrete unitary operator},
  journal = {Physical Review Letters},
  year    = {1994},
  volume  = {73},
  number  = {1},
  pages   = {58--61},
  doi     = {10.1103/PhysRevLett.73.58}
}

@article{Clements2016,
  author  = {Clements, W. R. and Humphreys, P. C. and Metcalf, B. J. and Kolthammer, W. S. and Walmsley, I. A.},
  title   = {Optimal design for universal multiport interferometers},
  journal = {Optica},
  year    = {2016},
  volume  = {3},
  number  = {12},
  pages   = {1460--1465},
  doi     = {10.1364/OPTICA.3.001460}
}

@article{zhang2021optical,
  title={An optical neural chip for implementing complex-valued neural network},
  author={Zhang, Hui and Gu, Mile and Jiang, XD and Thompson, Jayne and Cai, Hong and Paesani, Stefano and Santagati, Raffaele and Laing, Anthony and Zhang, Y and Yung, Man-Hong and others},
  journal={Nature communications},
  volume={12},
  number={1},
  pages={457},
  year={2021},
  publisher={Nature Publishing Group UK London}
}

@article{shen2017deep,
  title={Deep learning with coherent nanophotonic circuits},
  author={Shen, Yichen and Harris, Nicholas C and Skirlo, Scott and Prabhu, Mihika and Baehr-Jones, Tom and Hochberg, Michael and Sun, Xin and Zhao, Shijie and Larochelle, Hugo and Englund, Dirk and others},
  journal={Nature photonics},
  volume={11},
  number={7},
  pages={441--446},
  year={2017},
  publisher={Nature Publishing Group UK London}
}

@article{SinglePhoton2016,
  author  = {Ding, X. and He, Y. and Duan, Z.-C. and Gregersen, N. and Chen, M.-C. and Unsleber, S. and others},
  title   = {On-demand single photons with high extraction efficiency and near-unity indistinguishability from a resonantly driven quantum dot in a micropillar},
  journal = {Physical Review Letters},
  year    = {2016},
  volume  = {116},
  number  = {2},
  pages   = {020401}
}

@article{madsen2022quantum,
  title={Quantum computational advantage with a programmable photonic processor},
  author={Madsen, Lars S and Laudenbach, Fabian and Askarani, Mohsen Falamarzi and Rortais, Fabien and Vincent, Trevor and Bulmer, Jacob FF and Miatto, Filippo M and Neuhaus, Leonhard and Helt, Lukas G and Collins, Matthew J and others},
  journal={Nature},
  volume={606},
  number={7912},
  pages={75--81},
  year={2022},
  publisher={Nature Publishing Group UK London}
}

@article{larsen2025integrated,
  title={Integrated photonic source of {G}ottesman--{K}itaev--{P}reskill qubits},
  author={Larsen, Mikkel V and Bourassa, J Eli and Kocsis, Sacha and Tasker, Joel F and Chadwick, Robert S and Gonz{\'a}lez-Arciniegas, Carlos and Hastrup, Jacob and Lopetegui-Gonz{\'a}lez, Carlos E and Miatto, Filippo M and Motamedi, A and others},
  journal={Nature},
  volume={642},
  number={8068},
  pages={587--591},
  year={2025},
  publisher={Nature Publishing Group UK London}
}

@article{maring2024versatile,
  title={A versatile single-photon-based quantum computing platform},
  author={Maring, Nicolas and Fyrillas, Andreas and Pont, Mathias and Ivanov, Edouard and Stepanov, Petr and Margaria, Nico and Hease, William and Pishchagin, Anton and Lema{\^\i}tre, Aristide and Sagnes, Isabelle and others},
  journal={Nature Photonics},
  volume={18},
  number={6},
  pages={603--609},
  year={2024},
  publisher={Nature Publishing Group UK London}
}

@article{Thiele2020,
  author  = {Thiele, F. and vom Bruch, F. and Quiring, V. and Ricken, R. and Herrmann, H. and Eigner, C. and others},
  title   = {Cryogenic electro-optic polarisation conversion in titanium in-diffused lithium niobate waveguides},
  journal = {Optics Express},
  year    = {2020},
  volume  = {28},
  number  = {20},
  pages   = {28961--28968}
}

@article{bose2024anneal,
  title={Anneal-free ultra-low loss silicon nitride integrated photonics},
  author={Bose, Debapam and Harrington, Mark W and Isichenko, Andrei and Liu, Kaikai and Wang, Jiawei and Chauhan, Nitesh and Newman, Zachary L and Blumenthal, Daniel J},
  journal={Light: Science \& Applications},
  volume={13},
  number={1},
  pages={156},
  year={2024},
  publisher={Nature Publishing Group UK London}
}

@article{huang2006effect,
  title={Effect of deposition conditions on mechanical properties of low-temperature PECVD silicon nitride films},
  author={Huang, Han and Winchester, KJ and Suvorova, Alexandra and Lawn, BR and Liu, Yinong and Hu, XZ and Dell, JM and Faraone, Lorenzo},
  journal={Materials Science and Engineering: A},
  volume={435},
  pages={453--459},
  year={2006},
  publisher={Elsevier}
}

@article{reed2010silicon,
  title={Silicon optical modulators},
  author={Reed, Graham T and Mashanovich, Goran and Gardes, F Yand and Thomson, DJhttps},
  journal={Nature photonics},
  volume={4},
  number={8},
  pages={518--526},
  year={2010},
  publisher={Nature Publishing Group UK London}
}

@article{he2019high,
  title={{High-performance hybrid silicon and lithium niobate Mach--Zehnder modulators for \SI{100}{Gbit\per\second} and beyond}},
  author={He, Mingbo and Xu, Mengyue and Ren, Yuxuan and Jian, Jian and Ruan, Ziliang and Xu, Yongsheng and Gao, Shengqian and Sun, Shihao and Wen, Xueqin and Zhou, Lidan and others},
  journal={Nature photonics},
  volume={13},
  number={5},
  pages={359--364},
  year={2019},
  publisher={Nature Publishing Group UK London}
}

@article{yang2018characteristic,
  title={Characteristic study of silicon nitride films deposited by LPCVD and PECVD},
  author={Yang, Chris and Pham, John},
  journal={Silicon},
  volume={10},
  number={6},
  pages={2561--2567},
  year={2018},
  publisher={Springer}
}

@article{Chang2024,
author  = {Chang, Chang and Hou, Xiaoyu and Huang, Weixiong and Zhang, Aoxue and Sun, Yuhan and Li, Haolan and Shen, Li and Xu, Xiaochuan and Zou, Yi},
title   = {Lithium niobate modulator from room temperature to cryogenic conditions},
journal = {Optics Express},
year    = {2024}
}

@article{SilvestreACSNano2013,
author = {Silvestre, Ive and de Morais, Evandro A. and Melo, Angelica O. and Campos, Leonardo C. and Goncalves, Alem-Mar B. and Cadore, Alisson R. and Ferlauto, Andre S. and Chacham, Helio and Mazzoni, Mario S. C. and Lacerda, Rodrigo G.},
title = {Asymmetric Effect of Oxygen Adsorption on Electron and Hole Mobilities in Bilayer Graphene: Long- and Short-Range Scattering Mechanisms},
journal = {ACS Nano},
volume = {7},
number = {8},
pages = {6597-6604},
year = {2013},
doi = {10.1021/nn402653b},
note ={PMID: 23859671},
URL = {https://doi.org/10.1021/nn402653b},
eprint = {https://doi.org/10.1021/nn402653b}
}

@article{Mania2017,
doi = {10.1088/2053-1583/aa76f4},
url = {https://doi.org/10.1088/2053-1583/aa76f4},
year = {2017},
month = {jul},
publisher = {IOP Publishing},
volume = {4},
number = {3},
pages = {031008},
author = {Mania, E and Alencar, A B and Cadore, A R and Carvalho, B R and Watanabe, K and Taniguchi, T and Neves, B R A and Chacham, H and Campos, L C},
title = {{Spontaneous doping on high quality talc-graphene-hBN van der Waals heterostructures}},
journal = {2D Materials},
}

@article{Cadore2016,
    author = {Cadore, A. R. and Mania, E. and de Morais, E. A. and Watanabe, K. and Taniguchi, T. and Lacerda, R. G. and Campos, L. C.},
    title = {{Metal-graphene heterojunction modulation via H$_2$ interaction}},
    journal = {Applied Physics Letters},
    volume = {109},
    number = {3},
    pages = {033109},
    year = {2016},
    month = {07},
    issn = {0003-6951},
    doi = {10.1063/1.4959560},
    url = {https://doi.org/10.1063/1.4959560},
    eprint = {https://pubs.aip.org/aip/apl/article-pdf/doi/10.1063/1.4959560/14484562/033109_1_online.pdf},
}

@article{Pereira2019,
doi = {10.1088/2053-1583/ab0b23},
url = {https://doi.org/10.1088/2053-1583/ab0b23},
year = {2019},
month = {mar},
publisher = {IOP Publishing},
volume = {6},
number = {2},
pages = {025037},
author = {Pereira, C L and Cadore, A R and Rezende, N P and Gadelha, A and Soares, E A and Chacham, H and Campos, L C and Lacerda, R G},
title = {Reversible doping of graphene field effect transistors by molecular hydrogen: the role of the metal/graphene interface},
journal = {2D Materials},
}

@article{Gispare2021,
author = {Di Gaspare, Alessandra and Pogna, Eva Arianna Aurelia and Salemi, Luca and Balci, Osman and Cadore, Alisson Ronieri and Shinde, Sachin Maruti and Li, Lianhe and di Franco, Cinzia and Davies, Alexander Giles and Linfield, Edmund Harold and Ferrari, Andrea Carlo and Scamarcio, Gaetano and Vitiello, Miriam Serena},
title = {Tunable, Grating-Gated, Graphene-On-Polyimide Terahertz Modulators},
journal = {Advanced Functional Materials},
volume = {31},
number = {10},
pages = {2008039},
doi = {https://doi.org/10.1002/adfm.202008039},
url = {https://advanced.onlinelibrary.wiley.com/doi/abs/10.1002/adfm.202008039},
eprint = {https://advanced.onlinelibrary.wiley.com/doi/pdf/10.1002/adfm.202008039},
year = {2021}
}

@article{Viti2021,
author = {Viti, Leonardo and Cadore, Alisson R. and Yang, Xinxin and Vorobiev, Andrei and Muench, Jakob E. and Watanabe, Kenji and Taniguchi, Takashi and Stake, Jan and Ferrari, Andrea C. and Vitiello, Miriam S.},
title = {Thermoelectric graphene photodetectors with sub-nanosecond response times at terahertz frequencies},
journal = {Nanophotonics},
volume = {10},
number = {1},
pages = {89-98},
doi = {https://doi.org/10.1515/nanoph-2020-0255},
url = {https://onlinelibrary.wiley.com/doi/abs/10.1515/nanoph-2020-0255},
eprint = {https://onlinelibrary.wiley.com/doi/pdf/10.1515/nanoph-2020-0255},
year = {2021}
}

@article{Purdie2018,
   author = {D. G. Purdie and N. M. Pugno and T. Taniguchi and K. Watanabe and A. C. Ferrari and A. Lombardo},
   doi = {10.1038/s41467-018-07558-3},
   isbn = {4146701807558},
   issn = {20411723},
   issue = {1},
   journal = {Nature Communications},
   pages = {5387},
   publisher = {Springer US},
   title = {Cleaning interfaces in layered materials heterostructures},
   volume = {9},
   url = {http://dx.doi.org/10.1038/s41467-018-07558-3},
   year = {2018}
}

@article{Banszerus2016,
   author = {Luca Banszerus and Michael Schmitz and Stephan Engels and Matthias Goldsche and Kenji Watanabe and Takashi Taniguchi and Bernd Beschoten and Christoph Stampfer},
   doi = {10.1021/acs.nanolett.5b04840},
   issn = {15306992},
   issue = {2},
   journal = {Nano Letters},
   pages = {1387-1391},
   pmid = {26761190},
   title = {Ballistic Transport Exceeding 28 {$\upmu$}m in {CVD} Grown Graphene},
   volume = {16},
   year = {2016}
}

@article{Qiao2007,
author  = {Qiao, H. and Zhang, H. and Xu, J. and Wang, X. and Liu, H.},
title   = {{Temperature dependence of photorefractive effect in reduced near-stoichiometric LiNbO$_3$ crystal}},
journal = {Optics Communications},
year    = {2007},
volume  = {276},
number  = {1},
pages   = {130--133}
}

@article{Eltes2020,
  author  = {Eltes, F. and Villarreal-Garcia, G. E. and Caimi, D. and others},
  title   = {An integrated optical modulator operating at cryogenic temperatures},
  journal = {Nature Materials},
  year    = {2020},
  volume  = {19},
  pages   = {1164--1168},
  doi     = {10.1038/s41563-020-0725-5}
}

@article{Lee2021,
  author  = {Lee, B. and Kim, B. and Freitas, A. and Mohanty, A. and Zhu, Y. and Bhatt, G. and others},
  title   = {High-performance integrated graphene electro-optic modulator at cryogenic temperature},
  journal = {Nanophotonics},
  year    = {2021},
  volume  = {10},
  number  = {1},
  pages   = {99--104},
  doi     = {10.1515/nanoph-2020-0363}
}

@article{Liu2023,
  author  = {Liu, P. and Wen, H. and Ren, L. and Shi, L. and Zhang, X.},
  title   = {$\chi^{(2)}$ nonlinear photonics in integrated microresonators},
  journal = {Frontiers of Optoelectronics},
  year    = {2023},
  volume  = {16},
  pages   = {18},
  doi     = {10.1007/s12200-023-00073-4}
}

@article{Liu2011,
  author  = {Liu, M. and Yin, X. and Ulin-Avila, E. and others},
  title   = {A graphene-based broadband optical modulator},
  journal = {Nature},
  year    = {2011},
  volume  = {474},
  pages   = {64--67},
  doi     = {10.1038/nature10067}
}

@article{Wu2024,
  author  = {Wu, C. and Xu, K. and Wang, Y. and Hu, Q. and Zhang, H. and Dai, D. and others},
  title   = {Graphene-based silicon photonic electro-absorption modulators and phase modulators},
  journal = {IEEE Journal of Selected Topics in Quantum Electronics},
  year    = {2024},
  volume  = {30},
  number  = {4},
  pages   = {3400311},
  doi     = {10.1109/JSTQE.2024.3411058}
}

@article{Shu2018,
  author  = {Shu, H. and Su, Z. and Huang, L. and others},
  title   = {Significantly high modulation efficiency of compact graphene modulator based on silicon waveguide},
  journal = {Scientific Reports},
  year    = {2018},
  volume  = {8},
  pages   = {991}
}

@book{Neamen2012,
  author    = {Neamen, Donald A.},
  title     = {Semiconductor Physics and Devices: Basic Principles},
  edition   = {4},
  publisher = {McGraw-Hill Education},
  address   = {New York},
  year      = {2012},
  isbn      = {978-0073529585}
}

@article{Fan2017,
  author  = {Fan, M. and Yang, H. and Zheng, P. and Hu, G. and Yun, B. and Cui, Y.},
  title   = {Multilayer graphene electro-absorption optical modulator based on double-stripe silicon nitride waveguide},
  journal = {Optics Express},
  year    = {2017},
  volume  = {25},
  number  = {18},
  pages   = {21619--21629},
  doi     = {10.1364/OE.25.021619}
}

@article{Ji2020,
  author  = {Ji, L. and Chen, W. and Gao, Y. and Xu, Y. and Wu, C. and Wang, X. and Zhang, Z.},
  title   = {Low-power electro–optic phase modulator based on multilayer graphene/silicon nitride waveguide},
  journal = {Chinese Physics B},
  year    = {2020},
  volume  = {29},
  number  = {8},
  pages   = {084207},
  doi     = {10.1088/1674-1056/ab943b}
}

@inproceedings{Tiberi2025,
  author    = {Tiberi, M. and Wen, C.},
  title     = {Design of mid-infrared graphene optical modulators and detectors with gigahertz bandwidth on suspended silicon waveguides},
  booktitle = {2025 IEEE Silicon Photonics Conference (SiPhotonics)},
  year      = {2025},
  address   = {London, UK},
  pages     = {1--2},
  doi       = {10.1109/SiPhotonics64386.2025.10985595}
}

@article{Mohsin2015,
  author  = {Mohsin, M. and Neumaier, D. and Schall, D. and Otto, M. and Kurz, H.},
  title   = {Experimental verification of electro-refractive phase modulation in graphene},
  journal = {Scientific Reports},
  year    = {2015},
  volume  = {5},
  pages   = {10967},
  doi     = {10.1038/srep10967}
}

@article{Watson2024,
  author  = {Watson, H. F. Y. and Ruocco, A. and Tiberi, M. and others},
  title   = {Graphene phase modulators operating in the transparency regime},
  journal = {ACS Nano},
  year    = {2024},
  volume  = {18},
  number  = {44},
  pages   = {30269--30282},
  doi     = {10.1021/acsnano.4c02292}
}

@article{Wang2008,
  author  = {Wang, Feng and Zhang, Y. B. and Tian, C. and Girit, C. and Zettl, A. and Crommie, M. and Shen, Y. R.},
  title   = {Gate-variable optical transitions in graphene},
  journal = {Science},
  year    = {2008},
  volume  = {320},
  number  = {5873},
  pages   = {206--209}
}

@article{Falkovsky2008,
  author  = {Falkovsky, L. A.},
  title   = {Optical properties of graphene},
  journal = {Journal of Physics: Conference Series},
  year    = {2008},
  volume  = {129},
  number  = {1},
  pages   = {012004},
  doi     = {10.1088/1742-6596/129/1/012004}
}

@misc{COMSOLGraphene2022,
  author       = {Chen, X. (Tom)},
  title        = {Modeling graphene in high-frequency electromagnetics},
  howpublished = {COMSOL Blog},
  year         = {2022},
  month        = {June 15},
  note         = {Available at: \url{https://www.comsol.com/blogs/modeling-graphene-in-high-frequency-electromagnetics} (Accessed: Jan 2026)}
}

@misc{ANSYSGrapheneConductivity,
  author       = {{ANSYS Optics}},
  title        = {Graphene surface conductivity material model},
  howpublished = {Online documentation},
  note         = {Available at: \url{https://optics.ansys.com/hc/en-us/articles/360042244874-Graphene-surface-conductivity-material-model} (Accessed: Jan 2026)}
}

@article{Xie2022,
  author  = {Xie, S. and Veilleux, S. and Dagenais, M.},
  title   = {On-Chip High Extinction Ratio Single-Stage {Mach-Zehnder} Interferometer Based on Multimode Interferometer},
  journal = {IEEE Photonics Journal},
  year    = {2022},
  volume  = {14},
  number  = {4},
  pages   = {2237906},
  doi     = {10.1109/JPHOT.2022.3183214}
}

@book{Rezende2022,
  author    = {Rezende, S. M.},
  title     = {Introduction to Electronic Materials and Devices},
  publisher = {Springer},
  year      = {2022},
  isbn      = {978-3-030-81771-8}
}

@article{Banszerus2015,
  author  = {Banszerus, Luca and Schmitz, Michael and Engels, Stephan and Dauber, Jan and Oellers, Martin and Haupt, Federica and Watanabe, Kenji and Taniguchi, Takashi and Beschoten, Bernd and Stampfer, Christoph},
  title   = {Ultrahigh-mobility graphene devices from chemical vapor deposition on reusable copper},
  journal = {Science Advances},
  year    = {2015},
  volume  = {1},
  number  = {6},
  pages   = {e1500222},
  doi     = {10.1126/sciadv.1500222}
}

@article{Beliaev2022,
  author  = {Beliaev, L. Y. and Shkondin, E. and Lavrinenko, A. V. and Takayama, O.},
  title   = {Optical, structural and compositional properties of silicon nitride films deposited by reactive radio-frequency sputtering, low pressure and plasma-enhanced chemical vapor deposition},
  journal = {Thin Solid Films},
  year    = {2022},
  volume  = {763},
  pages   = {139568},
  doi     = {10.1016/j.tsf.2022.139568}
}

@article{Malitson1965,
  author  = {Malitson, I. H.},
  title   = {Interspecimen comparison of the refractive index of fused silica},
  journal = {Journal of the Optical Society of America},
  year    = {1965},
  volume  = {55},
  number  = {10},
  pages   = {1205--1208},
  doi     = {10.1364/JOSA.55.001205}
}

@article{Yakubovsky2017,
  author  = {Yakubovsky, D. I. and Arsenin, A. V. and Stebunov, Y. V. and Fedyanin, D. Y. and Volkov, V. S.},
  title   = {Optical constants and structural properties of thin gold films},
  journal = {Optics Express},
  year    = {2017},
  volume  = {25},
  number  = {21},
  pages   = {25574--25587},
  doi     = {10.1364/OE.25.025574}
}

@article{Droscher2010,
  author  = {Dr{\"o}scher, S. and Roulleau, P. and Molitor, F. and Studerus, P. and Stampfer, C. and Ihn, T. and Ensslin, K.},
  title   = {Quantum capacitance and density of states of graphene},
  journal = {Applied Physics Letters},
  year    = {2010},
  volume  = {96},
  number  = {15},
  pages   = {152104},
  doi     = {10.1063/1.3373529}
}

@article{Xia2009,
  author  = {Xia, J. and Chen, F. and Li, J. and Tao, N.},
  title   = {Measurement of the quantum capacitance of graphene},
  journal = {Nature Nanotechnology},
  year    = {2009},
  volume  = {4},
  pages   = {505--509},
  doi     = {10.1038/nnano.2009.177}
}

@article{Giovannetti2008,
  author  = {Giovannetti, G. and Khomyakov, P. A. and Brocks, G. and Karpan, V. M. and van den Brink, J. and Kelly, P. J.},
  title   = {Doping graphene with metal contacts},
  journal = {Physical Review Letters},
  year    = {2008},
  volume  = {101},
  number  = {2},
  pages   = {026803},
  doi     = {10.1103/PhysRevLett.101.026803}
}

@article{Malitson1972Al2O3,
  author  = {Malitson, I. H. and Dodge, M. J.},
  title   = {Refractive index and birefringence of synthetic sapphire},
  journal = {Journal of the Optical Society of America},
  year    = {1972},
  volume  = {62},
  number  = {12},
  pages   = {1405},
  doi     = {10.1364/JOSA.62.001405}
}

@article{AlKuhaili2004HfO2,
  author  = {Al-Kuhaili, M. F.},
  title   = {Optical properties of hafnium oxide thin films and their application in energy-efficient windows},
  journal = {Optical Materials},
  year    = {2004},
  volume  = {27},
  number  = {3},
  pages   = {383--387},
  doi     = {10.1016/j.optmat.2004.03.007}
}

@article{Yan2022,
  author  = {Yan, Yiyi and Kilchytska, Valeriya and Wang, Bin and Faniel, Sebastien and Zeng, Yun and Raskin, Jean-Pierre and Flandre, Denis},
  title   = {Characterization of thin {A}l{$_2$}{O}{$_3$}/{S}i{O}{$_2$} dielectric stack for {CMOS} transistors},
  journal = {Microelectronic Engineering},
  year    = {2022},
  volume  = {254},
  pages   = {111708}
}

@article{Sorianello2015,
  author  = {Sorianello, Vito and Midrio, Michele and Romagnoli, Marco},
  title   = {Design optimization of single and double layer graphene phase modulators in {SOI}},
  journal = {Optics Express},
  year    = {2015},
  volume  = {23},
  number  = {5},
  pages   = {6478--6490}
}

@article{Heo2011,
  author  = {Heo, J. and Chung, H. J. and Lee, Sung-Hoon and Yang, H. and Seo, D. H. and Shin, J. K. and Chung, U.-In and Seo, S. and Hwang, E. H. and Das Sarma, S.},
  title   = {Non-monotonic temperature dependent transport in graphene grown by chemical vapor deposition},
  journal = {arXiv preprint arXiv:1009.2506},
  year    = {2011},
  eprint  = {1009.2506},
  archivePrefix = {arXiv},
  primaryClass  = {cond-mat.mes-hall}
}

@article{Zhang2021LNReview,
  title={Integrated lithium niobate electro-optic modulators: when performance meets scalability},
  author={Zhang, Mian and Wang, Cheng and Kharel, Prashanta and Zhu, Di and Lon{\v{c}}ar, Marko},
  journal={Optica},
  volume={8},
  number={5},
  pages={652--667},
  year={2021},
  publisher={Optica Publishing Group}
}

@article{Li2017SiliconTransmitter,
  title={Electronic--photonic convergence for silicon photonics transmitters beyond 100 {G}bps on--off keying},
  author={Li, Ke and Liu, Shenghao and Thomson, David J. and Zhang, Weiwei and Yan, Xingzhao and Meng, Fanfan and Littlejohns, Callum G. and Du, Han and Banakar, Mehdi and Ebert, Martin and Cao, Wei and Tran, Dehn and Chen, Bigeng and Shakoor, Abdul and Petropoulos, Periklis and Reed, Graham T.},
  journal={Optica},
  volume={4},
  number={8},
  pages={938--945},
  year={2017},
  publisher={Optica Publishing Group}
}

\clearpage
\appendix
\section*{\Large Supplementary Information}

\subsection*{S1. Ridge Waveguide Design}
\label{sec:S1}
A waveguide can support different propagation modes of guided light, which are commonly differentiated by their polarization. Theoretically, the transverse electric (TE) mode is defined by a vanishing longitudinal electric field component ($E_z = 0$), while the transverse magnetic (TM) mode corresponds to solutions with $H_z = 0$. In practical ridge waveguides, these ideal conditions are not strictly fulfilled, and the modes are more accurately described as quasi-TE and quasi-TM. Nevertheless, the conventional TE and TM nomenclature is retained throughout this work. 

Since the main objective of this work is the implementation of a phase modulator based on a Mach–Zehnder interferometer (MZI), single-mode operation of the waveguide prior to graphene integration is required to ensure precise phase control and to avoid intermodal interference between TE and TM components \cite{Xie2022}. 

\subsubsection*{S1.1 Waveguide Core Dimensions}

Single-mode operation in a ridge waveguide can be achieved by properly selecting the core dimensions. Since the supported guided modes depend on the relation between the optical wavelength and the waveguide dimensions, higher-order modes can be suppressed when the waveguide width is below their cutoff condition, allowing only the fundamental mode to remain guided.

\begin{figure}[h!]
    \centering
    \includegraphics[width=0.6\columnwidth]{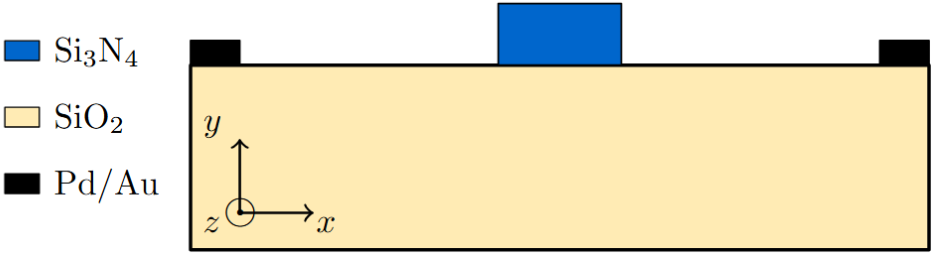}
    \caption{
    Cross-sectional schematic of RWG. The SiO$_2$ buried oxide (BOX) layer is \SI{2}{\um} thick and the upper cladding is air.
    }
    \label{fig:RWG_design}
\end{figure}

Therefore, the first step in the design is to determine the core width that ensures single-mode operation of the RWG. The cross-sectional geometry used in the simulations is illustrated in Fig.~\ref{fig:RWG_design}.

For the simulations, all materials were modeled as linear, isotropic, and non-magnetic media. Thus, the relative permittivity was defined using the relation $\varepsilon_r = n^2$ \cite{Rezende2022}, where $n$ is the refractive index of the material.

The refractive index values $n_{\mathrm{Si_3N_4}} = 1.9963$ \cite{Beliaev2022} and $n_{\mathrm{SiO_2}} = 1.4440$ \cite{Malitson1965} were taken at $\lambda = \SI{1.55}{\um}$ from experimental data reported in the literature. The refractive index of air was assumed to be $n=1$ at the same wavelength.

\begin{figure}[h!]
    \centering
    \includegraphics[width=0.6\columnwidth]{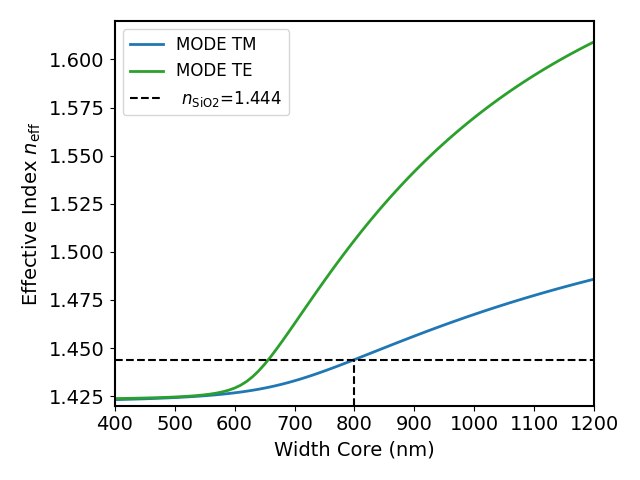}
    \caption{
    Effective index of the fundamental TE and TM modes as a function of the waveguide core width for a fixed core height of \SI{400}{\nm} at $\lambda = \SI{1.55}{\um}$.
    }
    \label{fig:Single_mode_operation}
\end{figure}

The simulation results are shown in Fig.~\ref{fig:Single_mode_operation} for a fixed core height of \SI{400}{\nm}. It can be observed that when the waveguide width is smaller than \SI{800}{\nm}, only the fundamental TE mode satisfies the guiding condition ($n_{\mathrm{eff}} > n_{\mathrm{SiO_2}}$). The TM fundamental mode and all higher-order modes are below cutoff and therefore cannot propagate.

Based on these results, the core dimensions were chosen as 400~nm in height and 800~nm in width. 

\subsubsection*{S1.2 Electrode--Waveguide Separation}

After establishing the single-mode operating region and the optimal core dimensions, an additional simulation was performed to evaluate the influence of the metal electrodes on the guided optical mode. In particular, the lateral separation between the waveguide core and the Au electrodes used to apply the bias voltage was varied in order to determine a distance that minimizes electrode-induced attenuation.

This simulation was carried out at $\lambda = \SI{1.55}{\um}$ for the TE mode. The refractive index of gold was taken as $\mathrm{Re}(n_{\mathrm{Au}})=0.52406$ and $\mathrm{Im}(n_{\mathrm{Au}})=10.742$ \cite{Yakubovsky2017}.

\begin{figure}[h!]
    \centering
    \includegraphics[width=0.6\columnwidth]{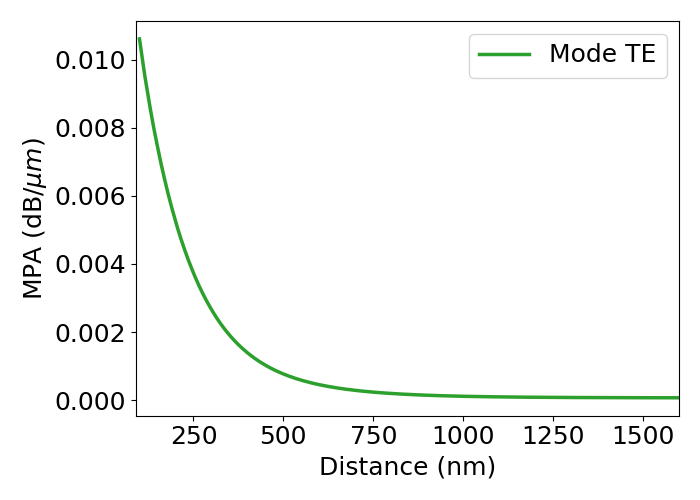}
    \caption{
    Mode Propagation Attenuation by Au Electrode for  $\lambda = \SI{1.55}{\um}$.}
    \label{fig:electrode_distance}
\end{figure}

From Fig.~\ref{fig:electrode_distance}, it can be observed that for electrode separations larger than \SI{1.2}{um} the propagation losses remain close to their minimum value. Therefore, the electrode edges should be positioned at least \SI{1.2}{um} from the waveguide core on each side during fabrication.

\end{document}